\documentclass[10pt,twocolumn]{article}

\usepackage{graphicx}
\usepackage{amsfonts}
\usepackage{amsmath}
\usepackage{amssymb}
\usepackage{mathrsfs} % cursive letters
\usepackage{lipsum}
\usepackage{color}
\usepackage{siunitx}
\usepackage{authblk}
\usepackage[margin=2cm]{geometry}
\usepackage{mathtools, cuted}

\usepackage{bm}
\usepackage{amsmath}

\title{Water-based peeling of thin hydrophobic films}

\author[1]{Sepideh Khodaparast}
\author[1,2]{Fran\c{c}ois Boulogne}
\author[2]{Christophe Poulard}
\author[1]{Howard A. Stone}
\affil[1]{Department of Mechanical and Aerospace Engineering, Princeton University, Princeton, NJ 08544, USA}
\affil[2]{Laboratoire de Physique des Solides, CNRS, Univ. Paris-Sud, Universit\'e Paris-Saclay, Orsay 91400, France}

\date{\today}
\begin{document}

\twocolumn[
    \begin{@twocolumnfalse}
        \maketitle
           \begin{abstract}
    Inks of permanent markers and water-proof cosmetics create elastic thin films upon application on a surface.
    Such adhesive materials are deliberately designed to exhibit water-repellent behavior.
    Therefore, patterns made up of these inks become resistant to moisture and cannot be cleaned by water after drying.
    However, we show that sufficiently slow dipping of such elastic films, which are adhered to a substrate, into a bath of pure water allows complete removal of the hydrophobic coatings.
    Upon dipping, the air-water interface in the bath forms a contact line on the substrate, which exerts a capillary-induced peeling force at the edge of the hydrophobic thin film.
    We highlight that this capillary peeling process is more effective at lower velocities of the air-liquid interface and lower viscosities.
    Capillary peeling not only removes such thin films from the substrate but also transfers them flawlessly onto the air-water interface.
           \end{abstract}
    \end{@twocolumnfalse}]

%%%%%%%%%%%%%%%%%%%%%%%%%%%%%%
%
% INTRODUCTION
%
%%%%%%%%%%%%%%%%%%%%%%%%%%%%%%

Water has always been a vital resource for life, not only for drinking, irrigation and industrial purposes \cite{Shannon:2008}, but also for its role as a natural cleaner \cite{Palabiyik:2015}.
However, recent developments in material science have lead to the invention of a large variety of water-repellent coatings \cite{Zhu:2006}, which are designed to feature relatively high adhesion to surfaces and resistance to moisture \cite{Ueda2013}.
These features are utilized in all manners of labeling and for creation of long-lasting patterns \cite{Kim:2011b}.
Nevertheless, in all of these applications there are circumstances where removal of such coatings is desired, and use of solvents must be avoided either for protecting the host substrate or for environmental  sustainability.
Well-known examples of such thin hydrophobic film materials are the inks of permanent markers \cite{Mammen:2006} and cosmetics \cite{Kim:2014} whose removal from a surface after drying requires special solvents or surfactants, which often cause damage to the substrate or irritation to the skin.
In this Letter, we discuss the effectiveness of a water meniscus for removing such thin films from substrates as the meniscus slowly propagates towards the films, pins at the edge of the film and eventually peels it off the substrate.

Peeling is an ubiquitous scientific problem that occurs repeatedly in diverse classical and modern engineering applications, which often involve cleaning procedures \cite{Boulogne2014b}, coating, manufacturing and transfer of thin films \cite{Cao:2008,Eda:2008,Lee:2011,Kim:2011a}, and fabrication of three-dimensional soft elastic structures \cite{Py:2007}.
The principles of peeling have been discussed in the pioneering works of Griffith, Obreimoff and Kendall as a quasi-static propagation of an interfacial crack between the laminating substance and the host substrate \cite{Griffith:1921,Obreimoff1930,Kendall:1971,Johnson:1971,Kendall:1975}.
According to the energy theory of fracture, the force required for peeling of a thin elastic film adhering to a solid substrate depends on the interfacial surface energy, the thickness and elasticity of the thin film, and the peeling angle \cite{Kendall:1971,Kendall:1975}.
Here we introduce \emph{capillary peeling} as an original mechanism for peeling of delicate waterproof materials.

The principle of our approach is to slowly dip a coated surface in a water bath.
We observed that under certain conditions the hydrophobic film detaches from the substrate and floats at the water interface.
To appreciate the basic steps of this procedure, we show snapshots of the peeling performed for marks made by Sharpie\textsuperscript{\textregistered} markers in Fig.~\ref{fig:sharpie}a (see the video in Supplemental Material SM).
The ink of these permanent markers contains a terpene phenolic resin, such as SYLVARES\textsuperscript{\textregistered} TP 2040~\cite{Mammen:2006}, which in combination with a carrier solvent and a pigment, is commonly used to promote adhesion.
Similar to the marks of the permanent markers, the thin film that is created upon drying of a drop of terpene dissolved in isopropanol on a glass slide can be removed by capillary peeling (Fig.~\ref{fig:sharpie}b).
Paradoxically, the hydrophobicity of the waterproof marks, which is supposed to provide them resistance to water, allows for flawlessly detaching and transferring the materials onto the air-water interface (see the SM).

%%%%%%%%%%%%%%%%%%%%%%%%%%%%%%%%%%%%%%%%%%%%%%%%%%%%%%%%%%%%%%%%
 \begin{figure}[ht!]
     \includegraphics[width=\linewidth]{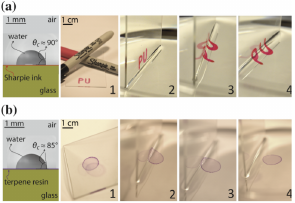}%
     \caption{Myth of the Sharpie\textsuperscript{\textregistered}.
     (a) The letters PU, written on a glass substrate with a red permanent Sharpie\textsuperscript{\textregistered} marker, are peeled off and transferred onto the air-water interface as the glass substrate is slowly dipped into a water bath. Patterns printed using Sharpie\textsuperscript{\textregistered} markers on glass are otherwise resistant to rinsing with water after drying. See the movies in the SM.
     (b) A similar phenomenon is observed for a dried droplet of terpene phenolic resin (SYLVARES\textsuperscript{\textregistered} TP 2040).
     For visualization, a small amount of Rhodamine B dye is added to the terpene phenolic resin.
     The advancing air-water interface moves at speed $U = 1$ $\mu$m/s.
     }
     \label{fig:sharpie}
 \end{figure}
 %%%%%%%%%%%%%%%%%%%%%%%%%%%%%%%%%%%%%%%%%%%%%%%%%%%%%%%%%%%%%%%%

 %%%%%%%%%%%%%%%%%%%%%%%%%%%%%%%%%%%%%%%%%%%%%%%%%%%%%%%%%%%%%%%%
\begin{figure}[ht!]
    \includegraphics[width=\linewidth]{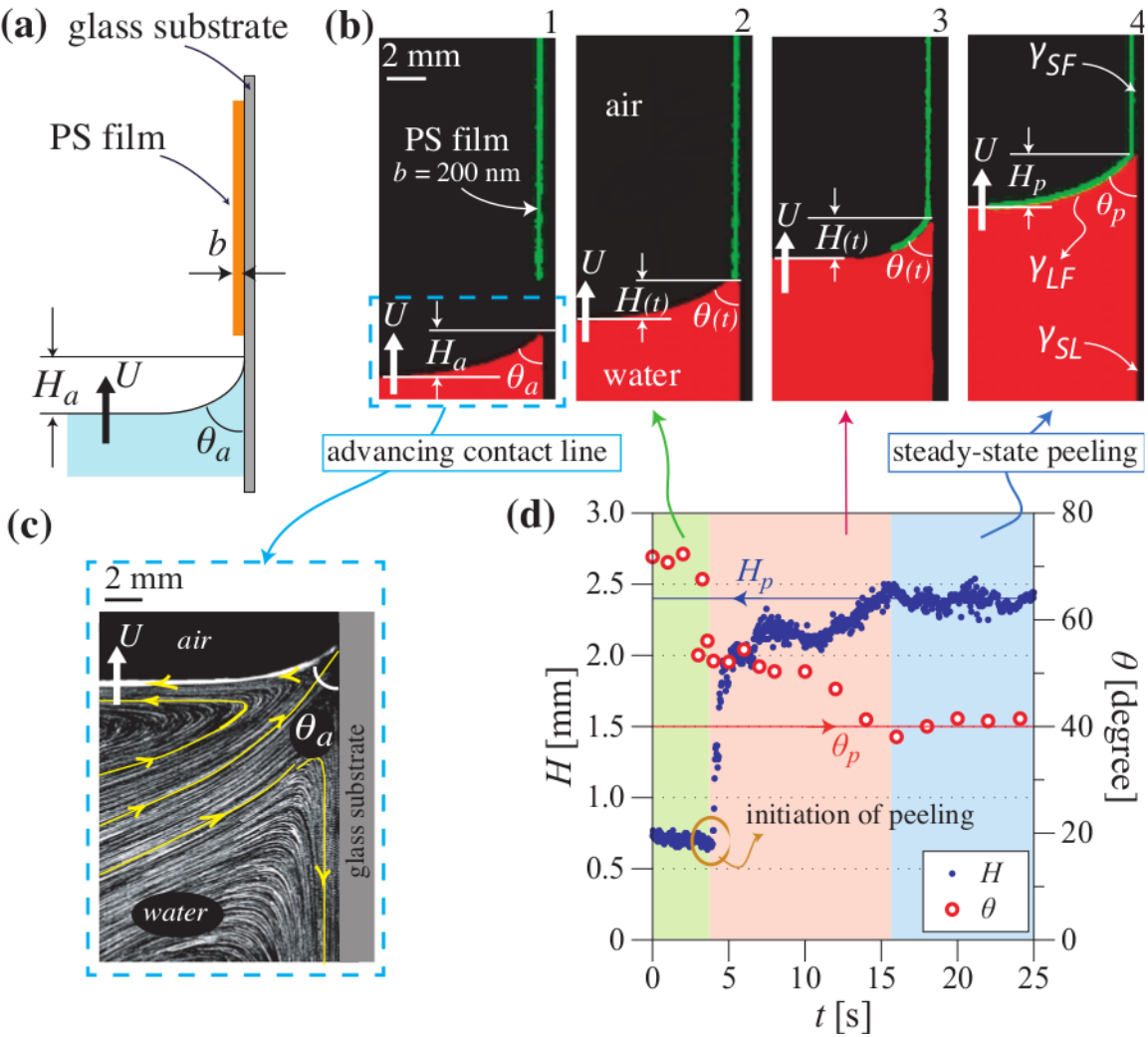}%
    \caption{Understanding capillary-driven peeling. (a) Schematic of the configuration used in the capillary peeling experiments. (b) Experimental visualizations of the main steps of the capillary peeling process.
    The liquid phase is labeled with a red fluorescent dye and the thin PS film ($b = 200$ nm) is covered with polystyrene microparticles to reflect the green light of the laser used in the experiments.
    The air-water interface moves with speed $U = 1$ $\mu$m/s on the substrate.
    The contact angle $\theta(t)$ changes from the advancing value $\theta_a$ on the clean substrate to the $\theta_p$ as the crack propagates in a steady-state manner.
    (c) On the hydrophilic glass substrate, water exhibits a split-injection flow pattern with a stagnation point at the contact point. (d) The time evolution of the experimental measurements of the height $H$ and the angle $\theta$ of liquid wedge confirms the distinct regimes present in a successful capillary peeling process.}
    \label{fig:intro_sch}
\end{figure}
%%%%%%%%%%%%%%%%%%%%%%%%%%%%%%%%%%%%%%%%%%%%%%%%%%%%%%%%%%%%%%%%

To discuss the criteria for initiation and propagation of effective capillary peeling of thin films, we examine the efficiency of this method in model experiments.
Indeed, as it is difficult to prepare films of terpene with uniform thickness and with annealed prestrain, we use polystyrene (PS) as a model hydrophobic material.
The peeling of PS has been used for different applications by several authors, \textit{e.g.} \cite{Bodiguel:2006,Huang:2007, Baumchen:2012}.
We prepared thin elastic PS films of well-controlled thicknesses $b \in [50,1200]$ nm adhered to clean glass substrates.
In all experiments, we use pure water and air as the liquid and the gas phase, respectively unless otherwise noted.
We perform the experiments by keeping the vertical substrate stationary, whilst moving the air-water interface upward using a motorized linear stage (Fig.~\ref{fig:intro_sch}a).
For visualization purposes, either red fluorescent dye (Fig.~\ref{fig:intro_sch}b) or seeding tracer particles (Fig.~\ref{fig:intro_sch}c) are added to the liquid phase.

%%%%%%%%%%%%%%%%%%%%%%%%%%%%%%%%%%%%%%%%%%%%%%%%%%%%%%%%%%%%%%%%

As the moving contact line reaches the edge of the thin PS film, which is adhered to the solid substrate, the surface tension applies a detachment force to the film, which drives the capillary peeling.
Fig.~\ref{fig:intro_sch}b shows the main steps of a successful capillary peeling process in a vertical configuration.
As the water meniscus moves upward at an average speed $U$, relative to the stationary substrate (Fig.~\ref{fig:intro_sch}b-1), the meniscus is pinned at the edge of the film (Fig.~\ref{fig:intro_sch}b-2).
At this point the water level far from the substrate continues to rise, whilst the meniscus remains pinned at the edge.
Finally, in case of successful peeling, defect-free thin films are peeled off of the vertical glass substrates and float on the free horizontal air-water interface (Fig.~\ref{fig:intro_sch}b-3,4).
On a hydrophilic substrate, water exhibits a split-injection flow pattern with a stagnation point at the three-phase contact point (Fig.~\ref{fig:intro_sch}c).
In this configuration, the air-liquid interface moves away from the substrate; if the contact line gets pinned when it reaches a hydrophobic film so that an interfacial crack between the film and the substrate is initiated, the flow profile guarantees that the film can be carried away from the substrate.

We show the time evolution of the height $H(t)$ and the angle $\theta(t)$ of the liquid wedge in Fig.~\ref{fig:intro_sch}d.
Initially, once the contact line is pinned at the film edge, the height $H(t)$ decreases from the capillary length $\ell_c = \sqrt{\frac{\gamma}{\rho g}} \approx 3$~mm, where $\gamma$ is the surface tension, $\rho$ is the density and $g$ is the gravitational acceleration, to a height of about 0.6 mm.
When the interfacial crack between the PS film and the substrate propagates, eventually the height of the liquid wedge $H$ and the associated macroscopic angle $\theta$ reach equilibrium values $H_p$ (blue line in Fig.~\ref{fig:intro_sch}d) and $\theta_p$ (red line in Fig.~\ref{fig:intro_sch}d), respectively.

Therefore, three main steps compose successful capillary peeling: the advancing motion of the contact line on a clean substrate, the initiation of the crack with the meniscus pinned to the edge of the film, and the steady-state propagation of the crack (Fig.~\ref{fig:intro_sch}d).
The interfacial energies of the materials, the velocity of the air-liquid interface, and the thickness of the film play crucial roles in the determining the effectiveness of the capillary peeling.
Next, we discuss the effects of these three parameters, to explain the operating conditions for the initiation of the capillary peeling and the steady-state propagation of the interfacial crack when the peeling is successful.
%%%%%%%%%%% RESULTS AND DISCUSSIONS %%%%%%%%%%%%%%%%%%%%%

For an ideal system without dissipation, the condition for initiating a crack between a thin elastic sheet and a smooth substrate is $G = W$, where $G$ is the effective mechanical energy/area to separate the interfaces~\cite{Kendall:1975} and $W$ is the thermodynamic work of adhesion, which is obtained from the interfacial energy balance to separate the film from the surface: the system gains the interfacial energy (per unit area) $\gamma_{SF}$ between the substrate and the film but loses the interfacial energies $\gamma_{SL}$ at the substrate-liquid interface and $\gamma_{LF}$ at the liquid-film interface (Fig. \ref{fig:intro_sch}b).
Therefore, the thermodynamic work of adhesion is
\begin{equation}\label{eq_W}
    W = \gamma_{SL}+\gamma_{LF}  - \gamma_{SF}.
\end{equation}

We can assess the relative importance of elastic effects on the peeling.
Assuming that the beam bends similar to a thin beam fixed at the tip of the crack, the gain of energy per unit width due to the elastic bending of the film is ${\cal E}_B \approx \frac{E w b^3}{12}\int_{0}^{\ell} \kappa^2 ds$, with $\kappa$ and $\ell$ being the curvature and length of the bent section of the film, respectively. Therefore,  considering the capillary length $\ell_c$ as the typical length scale in our geometry, the ratio of ${\cal E}_B$ to the gain of interfacial energy per unit width ${\cal E}_\gamma$ is typically on the order of $10^{-5}-10^{-1}$ for the range of thicknesses considered in our experiments.
Hence the contribution of the bending energy in Eq. (1) is neglected. The bending energy is expected to become considerable as the film thickness reaches around 10 $\mu$m in our experiments.

We use the Young-Dupr\'e equation expressed at the tip of the peeling front to relate the contact  angle observed for peeling, $\theta_p$, to these interfacial tensions.
Then, the thermodynamic work $W$ can be expressed as \cite{deGennes:2004}
\begin{equation}
    W=\gamma_{LF}(1-\cos\theta_p).
    \label{eq:Young}
\end{equation}
Using the contact angle value $\theta_p=40^\circ$ measured in Fig. \ref{fig:intro_sch}, and the value for the liquid-film interfacial energy $\gamma_{LF}=11$~mN/m determined experimentally \cite{Owens1969}, the thermodynamic work of adhesion for PS on glass is $W \approx 2.5$~mN/m. See Supplemental Material for brief description of the interfacial energy measurements.

This value can be compared to classical adhesion tests such as cleavage and blister tests with details presented in the SM, which yield $G \approx 17$~mN/m \cite{Kendall1994,Boulogne2017b}.
The difference between $G$ and $W$ indicates that energy dissipation occurs and that the condition for separation should be generalized as
\begin{equation}\label{eq:Kendall2}
    G-W = \sum_i D_i,
\end{equation}
where $D_i$ represent frictional, viscous or other dissipative forces (per unit length).

\begin{figure}[ht!]
     \center
     \includegraphics[width=\linewidth]{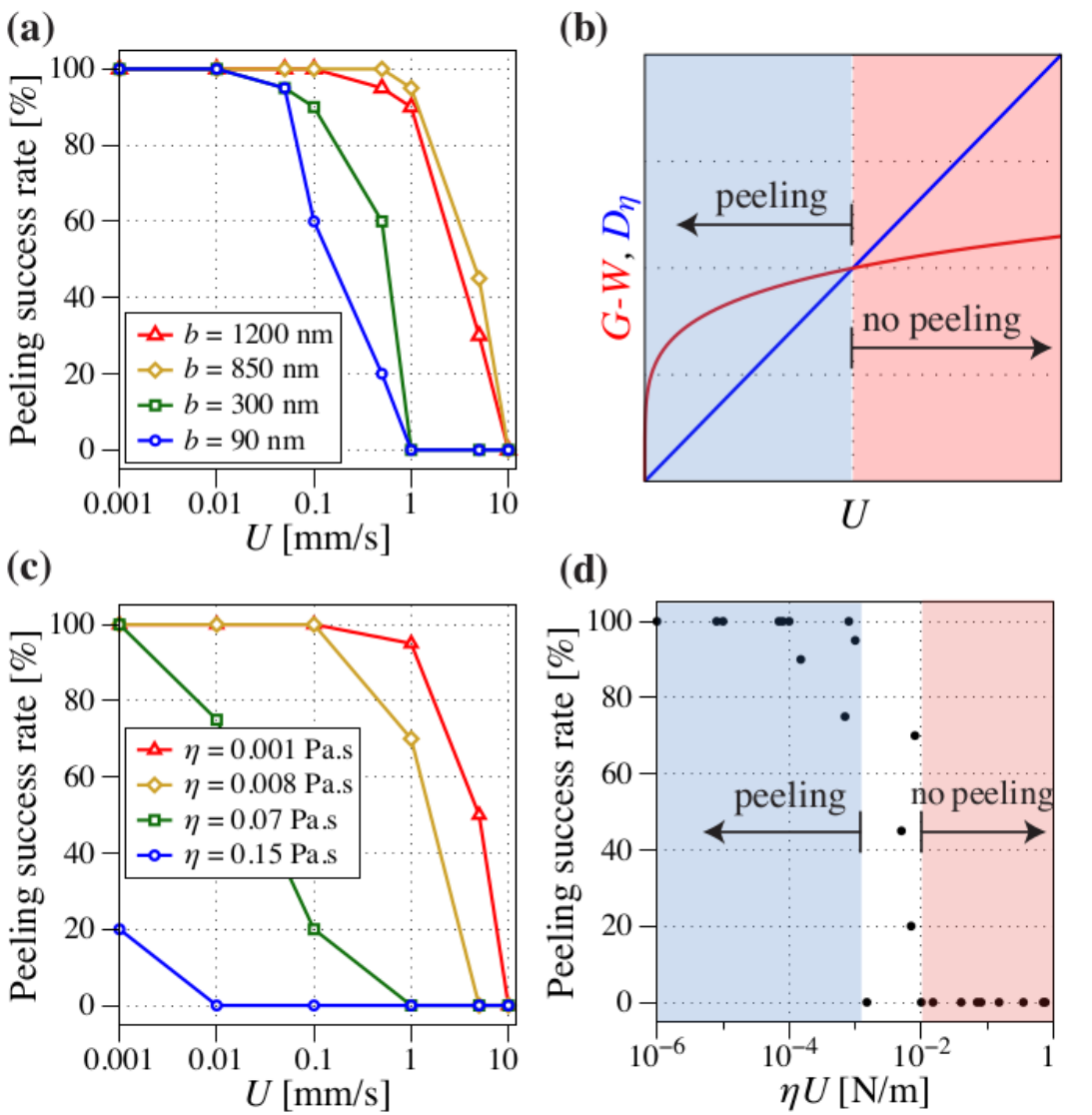}%
     \caption{Role of viscous dissipation $D_{\eta}$. (a) Capillary peeling becomes less efficient at higher speeds $U > 1$ mm/s, especially for very thin films $b < 100$ nm. The success rate is calculated for capillary peeling performed on 20 samples. (b) Velocity dependence of the adhesion energy $G$ and the viscous dissipation $D_\eta$. The blue domain corresponds to an energetically favorable peeling. (c) Capillary peeling becomes less efficient at higher viscosities $\eta \approx 10^{-2}$ Pa$\cdot$s for $b= 850$ nm. (d) The data from panel (c) collapse when plotted as a function of $\eta U$.
     Capillary peeling tends to fail for $\eta U > 10^{-2}$ N/m.}
     \label{fig:results1}
 \end{figure}

It is not straightforward to determine the dissipation mechanisms that occur when the triple line of the meniscus reaches the edge of the thin film, as in Fig.~\ref{fig:sharpie} and Fig.~\ref{fig:intro_sch}.
Experimentally, we report that the success rate of the peeling depends significantly on the speed of the contact line $U$ and also on the thickness of the film $b$ (Fig.~\ref{fig:results1}a).
There are at least two paths for energy dissipation: (i) viscous dissipation due to the fluid motion and (ii) the interfacial crack propagation, which induces mechanical dissipation that is a combination of modes I and II, respectively, for the cleavage and the shear of the film \cite{Hutchinson1991}.

%%%%%%%%%%%%%%%%%%%%%%%%%%%%%%%%%%%%%%%%%%%%%%%%%%%%%%%%%%%%%%%%
 %%%%%%%%%%%%%%%%%%%%%%%%%%%%%%%%%%%%%%%%%%%%%%%%%%%%%%%%%%%%%%%%
We can compare an estimate for the viscous dissipative forces/length $D_\eta$ of the liquid to pass over the film, which is proportional to $\eta U$ \cite{Guyon2001}, with the adhesion energy $G$ of the substrate-film interface.
It has been observed experimentally that the velocity dependence of the adhesion energy $G(U)$ is described by a power law, $G\propto U^\alpha$ with $\alpha$ an exponent smaller than one, which varies typically between $0.1$ and $0.4$ in experimental and numerical studies \cite{Baljon482,deruelle1995adhesion,Kendall:1971}.
Therefore, if the peeling speed is greater than a critical value, the dissipated energy through viscous effects is larger than the cost of the adhesion energy (see Fig.~\ref{fig:results1}b).
In this case, the liquid will pass over the film and the film will not detach.
In contrast, below this critical value, the capillary effects can initiate the crack.
Thus, the smaller the speed of the meniscus, the higher the success rate of the peeling process for a given film thickness $b$, as reported in Fig.~\ref{fig:results1}b.
A similar trend is observed with the viscosity of the liquid:  the smaller the viscosity, the higher the success rate of the capillary peeling (fig. \ref{fig:results1}c).
The data for the successful capillary peeling collapse when plotted as a function of $\eta U$ and we note that the critical transition, peeling to no peeling, occurs where $\eta U / G = O(1)$ (Fig. \ref{fig:results1}d).
We also recognize that for the thinnest films we have studied, e.g.\ typically $b < 100$ nm, defects present at the edge of the film and the substrate influence the success rate of the peeling (see SM).

\begin{figure}[ht!]
     \includegraphics[width=\linewidth]{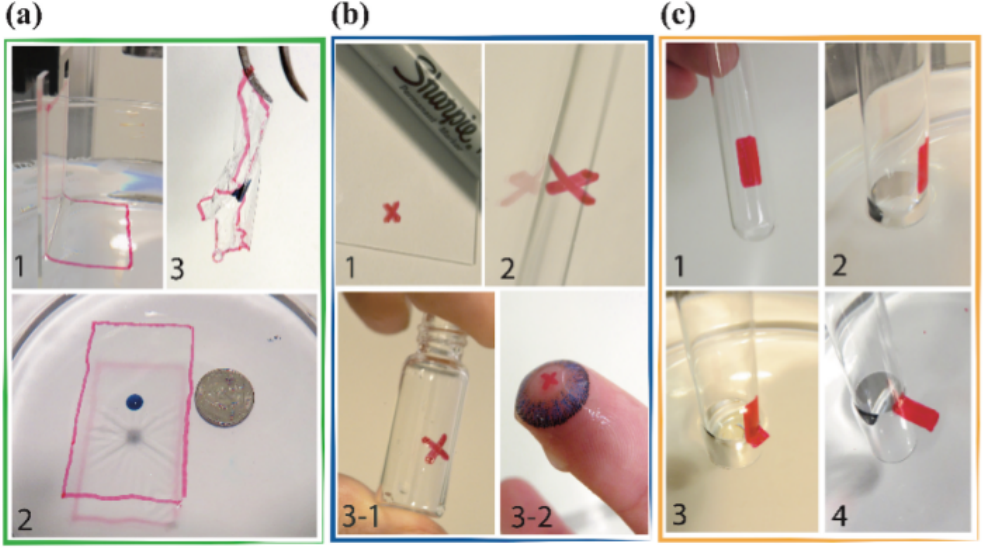}%
     \caption{Applications of capillary peeling.
     (a) Large-area thin films ($b = 200$ nm) can be (1) detached from the substrate, and (2) flawlessly transferred to the air-water interface.
     (3) Such thin films are otherwise difficult to handle or transferred to another substrate.
     (b) Complex patterns can be (1) detached from the original substrate, (2) moved onto the air-water interface and finally (3-1) and (3-2) be transferred onto unconventional substrates.
     Here marks made by Sharpie\textsuperscript{\textregistered} pens are transferred to the curved surface of a small glass bottle (3-1) and the delicate surface of a color contact lens (3-2).
     (c) Water-proof thin films can be removed from unconventional curved substrates.
     Here the mark made by a Sharpie\textsuperscript{\textregistered} pen is removed from the outer wall of a laboratory tube.\label{fig:application}}
 \end{figure}

Since capillary peeling is applied uniformly at the edge of the film, it allows peeling of defect-free large-area ultra thin films (Fig.~\ref{fig:application}a), which is of critical importance in fabrication of flexible and epidermal electronics \cite{Cao:2008,Eda:2008,Kim:2011b,Lee:2011,Kaltenbrunner:2013}, transparent deformable solar cells \cite{Kim:2015,Baca:2010}, and flexible screens \cite{Choi:2008,Kim:2011a}.
Moreover, the air-water interface can serve as a platform to transfer the detached hydrophobic marks onto unconventional receiver substrates by adding a dip-coating step to the process \cite{Scriven:1988} (Fig.~\ref{fig:application}b).
For this purpose, first the thin film is transferred onto the air-water interface. Next, the receiver substrate, that is initially submerged vertically in the water bath, is pulled out of the bath slowly.
Finally, the range of applications of capillary peeling is not limited to conventional flat substrates and provided that the criteria discussed above are met, this method can be used for peeling material from unconventional substrates of both positive and negative curvatures, following a similar procedure to that used for flat substrates (Fig.~\ref{fig:application}c).

In this Letter, we have shown that a water meniscus can effectively peel off thin elastic film from substrates to which they adhered.
In a counterintuitive manner, this water-based peeling method is more effective for detachment of hydrophobic material.
Providing that the capillary induced peeling force per unit length is larger than the adhesion energy per unit area of the thin film, different paths of energy dissipation may influence the effectiveness of this peeling method.
Among these, we demonstrated that viscous dissipation in the advancing liquid meniscus, which scales with $\mu U$, can significantly reduce the efficiency of the capillary peeling.
The fact that water naturally has high interfacial tension, relatively low viscosity and there exists a large variety of polymers and permanent inks that are designed to be water-proof (and thus hydrophobic), makes water one of the best candidates to be used as the liquid phase in the capillary peeling method.
The capillary peeling approach described here should prove applicable to a wide range of hydrophobic materials, printing techniques and emerging technologies.

\section*{Acknowledgments}
S.K. thanks the Swiss National Science Foundation for Early Mobility grant (P2ELP2-158896).
F.B. acknowledges that part of the research leading to these results received funding from the People Programme (Marie Curie Actions) of the European Union's Seventh Framework Programme (FP7/2007-2013) under REA grant agreement 623541.
H.A.S. thanks the NSF for support via grants CBET-1509347 and DMS 1614907.
We thank H. Hui for valuable discussions on the crack initiation and F. Restagno and T. Salez for discussions on the preparation of PS films.

        \bibliography{biblio}

        \bibliographystyle{unsrt}

\newpage
\clearpage

\section{Supplementary materials}

\subsection{Experiments}

{\bf{Preparation of substrates.}} Clean microscope glass slides (25 mm by 75 mm) were used in the peeling experiments.
Glass slides were left in a bath of acetone for 30 minutes.
Following this step they were thoroughly rinsed with DI water, and a solution of ethanol and acetone.
Glass slides were then dried with an air gun and heated at $100^{\circ}$C for 30 minutes prior to the experiments.

\begin{figure}[h]
     \center
     \includegraphics[width=\linewidth]{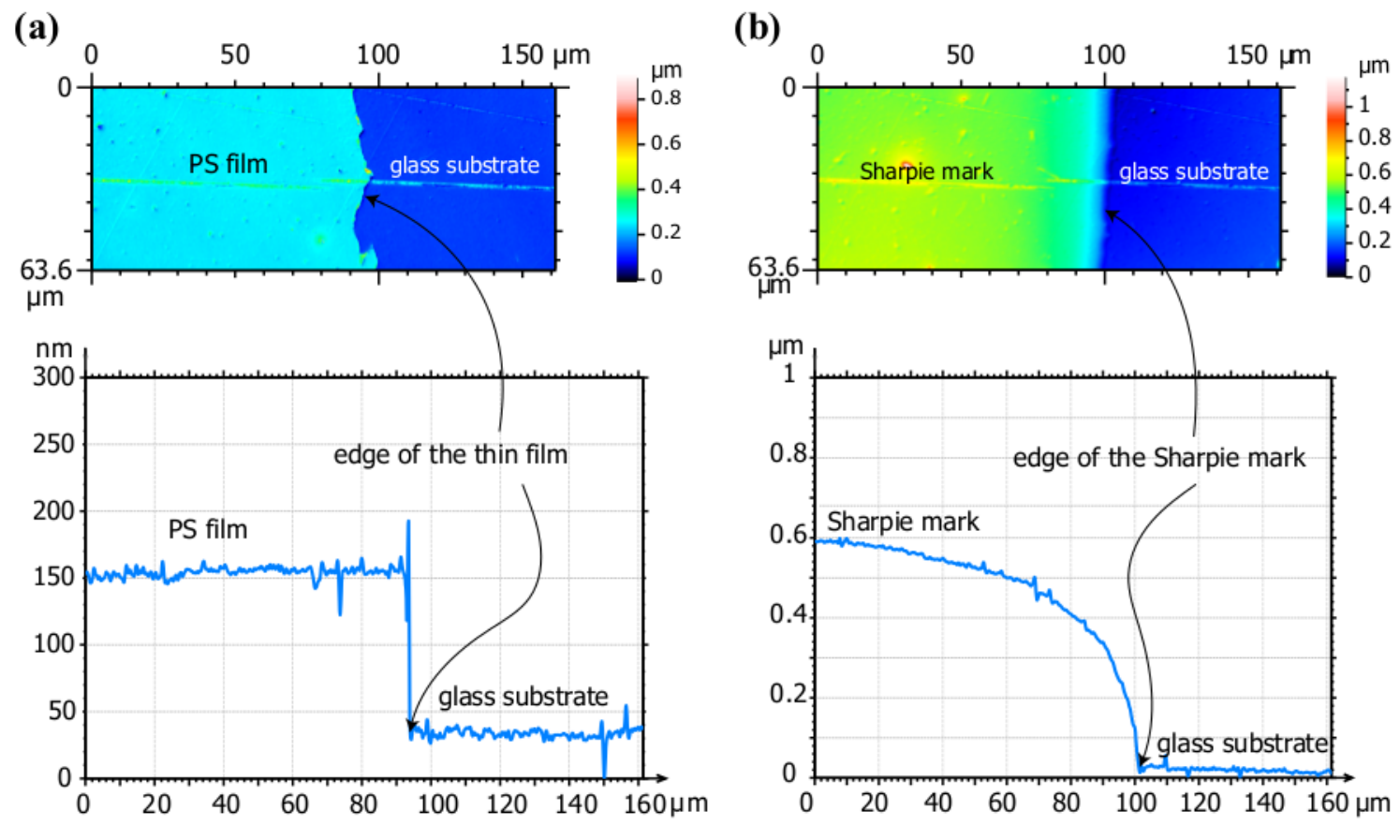}%
     \caption{Measurement of the film thickness performed by the optical profilometer.
     Top images are top views with a thickness color map and graphics are height profiles.
     (a) sample measurement performed on a PS film with an edge cut with a razor blade.
     (b) Sample measurement performed for a film made by a Sharpie\textsuperscript{\textregistered} marker.}
     \label{fig:FilmEdge}
 \end{figure}

{\bf{Preparation of thin films.}} The Sylvares\textsuperscript{\textregistered} TP 2040 sample used in the experiments was generously provided by Kraton Corporation.
Polystyrene PS films were prepared by spin coating a solution of polystyrene (Sigma-Aldrich, $M_w \simeq 280$ kg/mol) in toluene on the solid substrates at 2000 rpm for 30 seconds.
Different concentrations of PS were used to achieve film thicknesses ranging from $b = 50$ nm to $b = 2$ $\mu$m.A frame was then cut out with a sharp blade around the spin-coated films before they were annealed at $130^{\circ}$C under vacuum for two hours to release the possible pre-stress in the films.
This provided a sharp stepped edge at the boundaries of the film; see Fig.~\ref{fig:FilmEdge}.
Prior to the experiments, thin PS films were cooled down to room temperature under vacuum.
After annealing, the thickness of the polystyrene films and the marks from the Sharpie\textsuperscript{\textregistered} markers were measured by a Leica DCM 3D optical profilometer.
Spatial uniformity of the film thickness for polystyrene films was further confirmed using a Woollam M2000 Spectroscopic Ellipsometer.

In order to create a sharp edge, a frame was cut using a blade around the PS on the glass.
Optical profilometry measurements showed that this method provides a sharp edge, however, it can not provide a straight horizontal line at the edge of the film.
This method was used for all the samples used in the present experiments since it maintained the uniformity of the film thickness and resulted in a consistent shape profile at the edge of the film (Fig.~\ref{fig:FilmEdge}a).
In contrast, other methods such as dipping the edge of the film into a toluene solution was found not only to create random spatial profiles at the edge but also affects the quality and the thickness of the PS film close to the region in contact with the solvent.
We show a typical spatial profile of the thin films made by the Sharpie\textsuperscript{\textregistered} pens in Fig.~\ref{fig:FilmEdge}b.
As can be observed, the thickness of the mark is about 0.5 $\mu$m and a smooth profile is created at the edge after drying.

{\bf{Visualizations.}} Particle tracking and visualization  of the thin film peeling were performed on a thin sheet of light produced by a green laser module (ZM18B, Z-LASER).
For visualization purposes, the liquid phase is either seeded with tracer particles (PSP-20 from Dantec Dynamics) or red fluorescent dye (Rhodamine B, $M_w \simeq 480$ kg/mol).

{\bf{Capillary peeling experiments.}} For peeling experiments, the coated vertically oriented glass substrates were held stationary from the top.
The water bath was placed on a motorized linear translation stage (Thorlabs NRT series) underneath the substrate.
Experiments were performed by moving the water bath upward continuously at a given velocity 0.1 $\mu$m/s $< U <$ 10  mm/s.
For each experiment, capillary peeling is considered successful only if the entire film of 40 mm $\times$ 50 mm has been detached and transferred onto the air-water interface with no defects.

{\bf {Measurement of surface energies.}} The surface tension energies are measured with a Kr\"uss DSA-30S apparatus.
The PS-air surface tension $\gamma_{{\rm PS-air}} = 42.3$ mN/m is determined by using the Owens-Wendt-Rabel-Kaelbe model consisting in the contact angle measurement with six different liquids: water, glycerol, ethylene-glycol, hexadecane, octanol, diiodomethane \cite{Owens1969}.
Then, to determine the PS-water surface tension $\gamma_{{\rm PS-water}}$, we measure the water-air surface tension with a pendant drop technique.
From the contact angle measurement of a sessile drop deposited on PS, we calculate with the Young-Dupr\'e equation to find $\gamma_{{\rm PS-water}}=10.6$ mN/m.

\begin{figure}[h]
 \center
 \includegraphics[width=\linewidth]{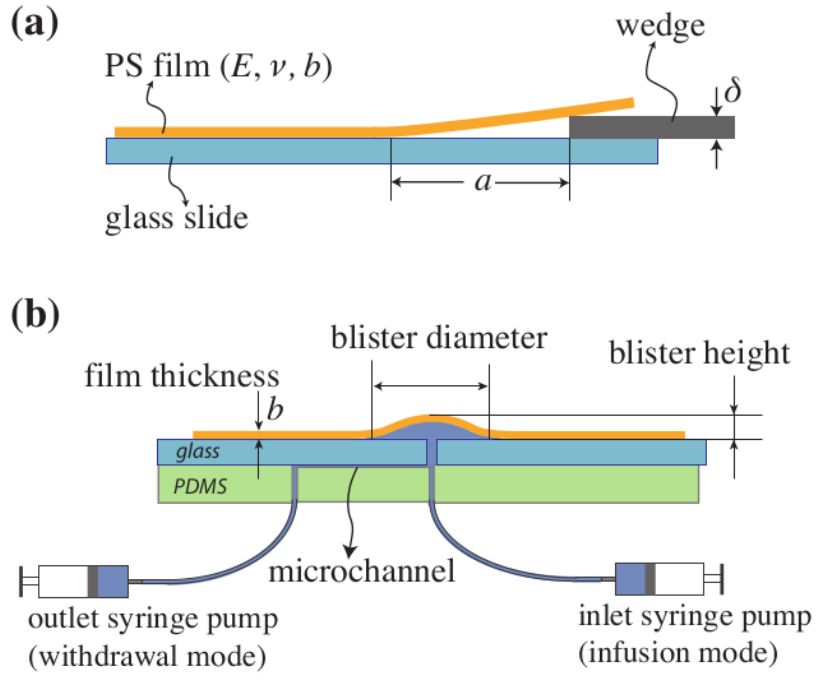}%
 \caption{Schematics of adhesion measurement experiments. (a) Cleavage test: a razor blade of thickness $\delta$ is used as a wedge to create an interfacial crack between the film and the substrate. (b) Blister test: the thin polystyrene film is mounted on a glass substrate which contains an injection opening in the center. Dyed water in injected through this opening and a blister is formed under the film. The diameter and the height of the blister were used for estimation of the adhesion energy. For more details please refer to \cite{Boulogne2017b}.}
 \label{fig:adhesion}
 \end{figure}

{\bf {Measurement of adhesion energy.}} Adhesion energies of PS films on glass substrates were measured with a cleavage test by propagating an interfacial crack between the film and the substrate with a wedge.
Measurements of the cleavage test were compared to those obtained using a blister test for PS on glass, as schematically presented in Fig.~\ref{fig:adhesion} \cite{Boulogne2017b}.
%As the wedge must be pushed between the film and the substrate, this test prevents us to measure the adhesion energy on thicknesses smaller than few tens of micrometers.
We prepared the polystyrene film with a solvent-free method by melting the polymer for the cleavage test.
Glass slides are prepared with the protocol described before.
Polystyrene pellets (Sigma-Aldrich, $M_w \simeq 280$ kg/mol, as used before) are placed between two glass slides, themselves placed between two aluminum plates (1 cm $\times$ 10 cm $\times$ 10 cm) held together by three screws.
This press is placed in a oven at an initial temperature of $130^\circ$C.
Screws are regularly tightened to squeeze the pellets and the temperature is raised until $220^\circ$C and the desired thickness is obtained.
The temperature is decreased to $130^\circ$C and maintained for 12 hours to relax the difference of thermal dilation between the film and the substrate and is then further decreased to the room temperature.
Aluminum plates and one of the glass slides are removed and the film is annealed again.
The film thickness is typically of $b = 100$ $\mu$m in our experiments.

A razor blade of thickness $\delta=385$ $\mu$m is used as a wedge.
The wedge is placed parallel and in contact with the substrate and pushed by a translational stage.
Once the crack is initiated, we stop the motor and measure the distance $a$ between the wedge and the crack tip; see Fig.~\ref{fig:adhesion}a.
From the minimization of the sum of the bending and adhesion energies, the energy release rate is \cite{Kendall1994}
\begin{equation}\label{eq:adhesion_cleavage}
    G = \frac{3}{16} \frac{E b^3 \delta^2}{(1-\nu^2) a^4},
\end{equation}
where $\delta$ is the height of the blade, $a$ the length of the crack and $\nu$ is the Poisson's ratio for polystyrene $\nu = 0.33$.

{\bf {Viscosity of the liquid phase.}} In order to investigate the effect of the working liquid,  water-glycerol solutions at different concentrations of glycerol were used.  Measurements presented in Fig. 3c and d of the letter were obtained by performing experiments with four aqueous solutions of glycerol with wt.$\%$ of 0, 60, 85 and 92 at room temperature $T = 25^{\circ }$C.

\subsection{Effective capillary peeling}

\begin{figure}[ht]
 \center
 \includegraphics[width=\linewidth]{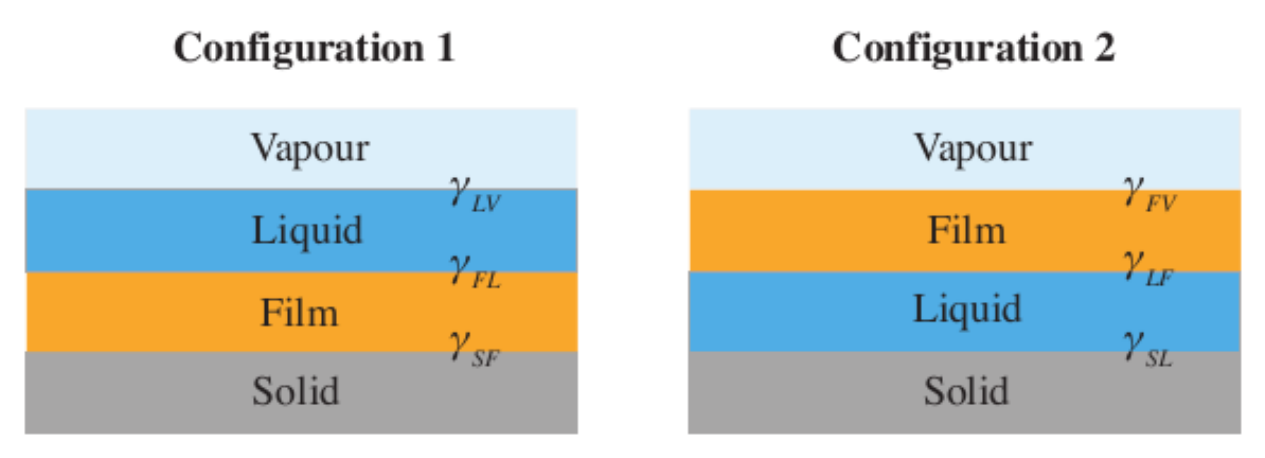}%
 \caption{Comparison of the two distinct configurations occurring when the film is not peeled and instead is wetted be the liquid (configuration 1) and when film is peeled by meniscus and is separated from the substrate by the liquid phase (configuration 2).}
 \label{fig:effective_peeling}
 \end{figure}

From a thermodynamical point of view, a successful peeling must satisfy two conditions.
First, configuration 2 (peeling) must be energetically more favorable than configuration 1 (no peeling) as shown schematically in Fig.~\ref{fig:effective_peeling}, \textit{i.e.}

\begin{equation}\label{SI:1}
\gamma_{SL} + \gamma_{LF} + \gamma_{FV} <  \gamma_{SF} + \gamma_{LF} + \gamma_{LV}.
\end{equation}

Second, as stated in the main paper, the penetration of the liquid in the crack tip must be energetically favorable, which is expressed as

\begin{equation}\label{SI:2}
W = \gamma_{SL} + \gamma_{LF}  - \gamma_{SF} > 0,
\end{equation}
where $W$ is defined as the thermodynamical work of adhesion.

By combining equations (\ref{SI:1}) and (\ref{SI:2}), we obtain

\begin{equation}\label{SI:3}
0 < W < S,
\end{equation}
where $S = \gamma_{LF} + \gamma_{LV} - \gamma_{FV}$ is the spreading parameter of the liquid on the film.
Therefore, equation (\ref{SI:3}) implies that the liquid must partially wet the film ($S>0$), which corresponds to our observations that a successful peeling is observed with hydrophobic films.
As detailed further in the main paper, this derivation assumes the absence of dissipations.
Taking into account those dissipations leads to the criterion
\begin{equation}\label{SI:4}
G < S.
\end{equation}

        \end{document}